\documentclass[aps,pra,twocolumn,showpacs,amsmath,amsfonts]{revtex4-1}

\usepackage{bm,units}

\newcommand{\deriv}[2]{\frac{\partial #1}{\partial #2}}
\newcommand{\ii}{\mathrm{i}}
\newcommand{\ee}{\mathrm{e}}
\newcommand{\dd}{\mathrm{d}}
\newcommand{\scatt}{a_{\mathrm{sc}}}
\newcommand{\pithree}{\pi^{\nicefrac{3}{2}}}
\newcommand{\pifive}{\pi^{\nicefrac{5}{2}}}
\newcommand{\scc}{\mathrm{sc}}

\begin{document}

\title{Variational methods with coupled Gaussian functions for Bose-Einstein 
condensates with long-range interactions. I. General concept}

\author{Stefan Rau}
\author{J\"org Main}
\author{G\"unter Wunner}
\affiliation{Institut f\"ur Theoretische Physik 1, Universit\"at Stuttgart,
  70550 Stuttgart, Germany}
\date{\today}

\begin{abstract}
The variational method of coupled Gaussian functions is applied to 
Bose-Einstein condensates with long-range interactions.
The time-dependence of the condensate is described by dynamical equations
for the variational parameters. 
We present the method and analytically derive the dynamical equations 
from the time-dependent Gross-Pitaevskii equation.
The stability of the solutions is investigated using methods of nonlinear 
dynamics.  The concept presented in this paper will be applied to 
Bose-Einstein condensates with monopolar $1/r$ and dipolar $1/r^3$ 
interaction in the subsequent paper [S.~Rau et al., Phys. Rev. A, submitted],
where we will present a wealth of new phenomena obtained by using 
the ansatz with coupled Gaussian functions. 
\end{abstract}


\maketitle

\section{Introduction}
\label{sec:intro}
The experimental realization of Bose-Einstein condensates (BEC) with 
$^{52} \mathrm{Cr}$ atoms \cite{EXGriesmaierPRL94,BeaufilsPRA77}, 
with a strong dipole-dipole interaction has given new impetus
to theoretical investigations of BEC with long-range interactions.

The theoretical description of Bose-Einstein condensates in the dilute limit
in the framework of the extended Gross-Pitaevskii equation (GPE) is well known. 
The derivation of the extended Gross-Pitaevskii equation from a many particle 
Schr\"odinger equation is part of many text books on quantum mechanics or 
Bose-Einstein condensates \cite{PitaevskiiBlackBook}. 
For a long-range interaction of the form 
$W_{\mathrm{lr}}(\bm r, \bm r') \propto |\bm r - \bm r'|^\alpha$,
the time-dependent GPE can be brought into particle number scaled 
dimensionless form using appropriate units (for monopolar or dipolar 
condensates, see \cite{PapadopoulosPRA76,gelbPatrick}), and reads
\begin{align}
\label{eq:GPE_general_timeindependent}
\Biggl[ &- \Delta 
 + \gamma_x^2 x^2 + \gamma_y^2 y^2 + \gamma_z^2 z^2 
 + 8 \pi \scatt \left|\psi(\bm r,t) \right|^2 \nonumber \\
&+ \int \! \dd^3 \bm{r'} \, W_\mathrm{lr}\!\left( \bm r, \bm r' \right)\left|
 \psi\left( \bm r',t \right) \right|^2 \Biggr]
 \psi(\bm r) = \ii\frac{\dd}{\dd t} \psi (\bm r,t) \; .
\end{align}
The terms in Eq.~\eqref{eq:GPE_general_timeindependent} describe the 
short-range contact interaction between two particles, the s-wave scattering, 
$V_\mathrm{sc}=8\pi\scatt|\psi(\bm r)|^2$, a harmonic model for external 
magnetic trapping of the condensate, 
$V_\mathrm{t}=\gamma_x^2 x^2 + \gamma_y^2 y^2 + \gamma_z^2 z^2$,
and long-range interactions between two particles
\begin{align*}
 V_\mathrm{lr}(\bm r) = \int \dd^3 {\bm r}' \, W_\mathrm{lr}(\bm r, \bm r')
 |\psi(\bm r')|^2 \; .
\end{align*}
The mean field energy reads
\begin{align}
\label{eq:mf_functional_theory}
 E_{\rm mf} = \left\langle \psi | -\Delta +  V_\mathrm{t}
 + ( V_\mathrm{sc} + V_\mathrm{lr})/2 | \psi  \right\rangle \; .
\end{align}
So far, in most publications one of two methods for solving the GPE is used. 
The ground state for the GPE with long-range interaction in 
Eq.~\eqref{eq:GPE_general_timeindependent} is obtained by minimizing the 
energy functional \eqref{eq:mf_functional_theory} using different approaches.

The first method are numerical lattice calculations \cite{Ronen06a}, 
either the minimization of the energy with conjugate gradients or 
imaginary time evolution of an initial wave function using the split-operator 
method and FFT.
The numerical calculations are on the one hand very accurate if they are 
carried out on sufficiently large grids, on the other hand, however, they 
may turn out laborious and may take a long computational time.

The second well-established method is using a simple variational ansatz.
A common ansatz is to assume a Gaussian type wave function 
\cite{ODellPRL84,holgerPRA78,PerezPRL,SantosPRL85,YiYouPhysRevA61}.
This technique allows to gain physical insight, as it often provides 
qualitatively correct, although quantitatively inaccurate results.
The extension and improvement of variational techniques for BECs with
long-range interactions is the major challenge of this paper. 

We propose, as a third approach, an improved variational ansatz with coupled 
Gaussian functions.
The method was originally proposed by Heller \cite{hellerJCP65,hellerJCP75}
to describe atomic and molecular quantum dynamics.
Using the ansatz
\begin{align}
 \psi(\bm{x},t) 
 =\sum_n\exp\{&\ii[(\bm x-\bm x_{t,n})\bm A_{t,n}(\bm x-\bm x_{t,n})\nonumber \\
 &+\bm p_{t,n}(\bm x-\bm x_{t,n})+\gamma_{t,n}]\}
\end{align}
where the symmetric matrix $\bm A_{t,n}$, the vector $\bm p_{t,n}$ and the
scalar $\gamma_{t,n}$ describe the width, momentum and weight of the Gaussian
wave packet, respectively, Heller approximated the dynamics of quantum wave
packets following classical trajectories.
The method was recently successfully applied to the dynamics of atoms in 
external fields \cite{FabcicI,FabcicII} using up to $N=100$ coupled functions.  

We apply this method to Bose-Einstein condensates with long-range interactions
using an ansatz with $N$ coupled Gaussian functions centered at the origin, 
$\bm x_{t,n}=0$, and with $\bm p_{t,n}=0$, viz.\
\begin{equation}
\label{def:dipolartrialfunction}
 \psi(\bm{r},t) =\sum_{k = 1}^{N} \left[ \ee^{\ii \gamma^{k}} 
 \prod_{\alpha} \left( \ee^{\ii a_\alpha^{k}x_\alpha^{2}} \right) \right] \; ,
\end{equation}
where $\alpha\in\{x,y,z\}$ for a BEC without symmetries,
$\alpha \in \{\varrho,z\}$ for an axisymmetric, and $\alpha = r$ for 
a spherically symmetric BEC.

The ansatz in Eq.~\eqref{def:dipolartrialfunction} is rather general and is 
not only able to describe condensates with spherical or axial symmetry,
but also nonsymmetric condensates in arbitrary trap geometries or even 
anisotropic solitons \cite{Tikhonenkov08a,ruedigerDA}.
The method is also well suited for long-range interactions because by its 
very nature it requires integrals over the entire wave function which
show a similar behavior as the integrals for local interactions.
Note that, as is typical for a Gaussian basis set, the individual Gaussian
functions in Eq.\ \eqref{def:dipolartrialfunction} are not orthogonal
amongst each other.
In this paper we present the theoretical concept of the method and derive the 
necessary equations for arbitrary long-range interaction.
For monopolar $(1/r)$ or dipolar $(1/r^3)$ interaction selected results
have already been presented in \cite{Rau10}.
Results elaborated in more detail are subject of the subsequent paper 
\cite{paper2}.
With the use of multiple Gaussians and thus an extended set of variational 
parameters, we are not only able to describe the stable ground state and 
the metastable stationary states, but can also identify the types of 
bifurcations where branches emerge or stability changes take place. 

The paper is organized as follows.
In Sec.~\ref{sec:lachlan_theory} we apply a time-dependent variational 
principle to the Gross-Pitaevskii equation 
\eqref{eq:GPE_general_timeindependent} and obtain dynamical equations, 
which describe the time-dependence of the variational parameters.
In Sec.~\ref{sec:dipolar_integrals} we evaluate the integrals that are 
needed to set up this nonlinear set of dynamical equations. 
In Sec.~\ref{sec:numericalprocedure} we present different methods 
for obtaining stationary solutions, and in Sec.~\ref{sec:theory:EW} we 
investigate the stability of those states.
Conclusions are given in Sec.~\ref{sec:conclusion}.

\section{Time Dependent Variational Principle}
\label{sec:lachlan_theory}
The first variational principle in order to optimize a variational wave 
function that comes to mind is minimizing the mean field energy functional. 
For few variational parameters the analytical calculation is done easily, 
but it gets increasingly difficult if more variational parameters are 
included in the wave function. 
For a BEC with long-range interaction and more than three parameters, it 
turns out to be almost impossible to calculate all derivations analytically. 
Therefore we will introduce a different approach based on the 
Dirac-Frenkel-McLachlan variational principle \cite{Lachlan,DiracTDVP}. 
The application of this variational principle yields a set of differential 
equations for the parameters $\bm z$ of the trial wave function 
$\psi(t)=\psi(\bm z(t))$, where 
\begin{equation}
 \bm{ z}(t) = \left(z_1(t),z_2(t),\hdots,z_M(t)\right)
\end{equation}
is the vector consisting of all $M$ variational parameters.
The time dependence of a quantum system is described by the respective 
Schr\"odinger or Gross-Pitaevskii equation
\begin{equation}
\label{eq:se}
 H \psi(t) = \ii \frac{\dd}{\dd t}\psi(t) \; .
\end{equation}
The variational principle of McLachlan minimizes the difference between the 
left and the right hand side of the respective Schr\"odinger equation 
\eqref{eq:se} with respect to the trial wave function,
\begin{align}
\label{eq:lachlan_theory}
 I &= || \ii \phi(t) - H \psi(t)||^2 \stackrel{!}{=} \min .
\end{align}
For any $t$, $\psi(t)$ is supposed to be fixed and given, and the quantity 
$I$ is minimized by varying $\phi$.
Afterwards $\phi$ is set equal to $\phi=\dot\psi$.
The time dependence of the trial wave function carries over to the time 
dependence of the variational parameters, $\psi(t)= \psi(\bm z(t))$.
We consider variations of $I$ in Eq.~\eqref{eq:lachlan_theory} with respect 
to $\phi$,
\begin{align}
\label{eq:deltaI2}
   \delta I &= \langle\delta \phi|\phi\rangle+\langle\phi|\delta\phi\rangle
  + \ii \langle\delta\phi|H\psi\rangle - \ii \langle H \psi|\delta\phi\rangle
 \nonumber \\
 &= \langle\delta \phi|\phi + \ii H \psi\rangle
  + \langle\phi-\ii H \psi|\delta\phi\rangle \; ,
\end{align}
where the variation of the time derivative of the wave function $\delta\phi$ 
carries over to variations of the parameters $\bm z$,
\begin{equation}
\label{eq:whyTheHellLachlanParameterFixed}
 |\delta \phi\rangle = |\delta \dot \psi (\bm z, \dot{\bm z})\rangle = 
 \left|\deriv{\dot \psi}{\bm z} \delta\bm z\right\rangle
 + \left|\deriv{\dot \psi}{\dot{\bm z}} \delta \dot{\bm z}\right\rangle \; .
\end{equation}
The first term vanishes, since we minimize $I$ under the condition that 
$\psi(t)$ is fixed and therefore that all parameters $\bm z$ are fixed as 
well, and we obtain
\begin{align}
\label{eq:deltaphi}
   |\delta \phi\rangle
 = \left|\deriv{}{\dot{\bm z}}\left(\deriv{\psi}{\bm z} \dot{\bm z}
   \right)\delta \dot{\bm z}\right> 
 = \left|\deriv{\psi}{\bm{ z}}\delta\dot{\bm z}\right>.
\end{align}
We insert Eq.~\eqref{eq:deltaphi} in Eq.~\eqref{eq:deltaI2}, set 
$\phi = \dot \psi$, and obtain as condition for the vanishing variation of $I$
\begin{equation}
\label{eq:braketcondif}
  \delta I = \left\langle\deriv{\psi}{\bm{ z}}\delta\dot{\bm z}\bigg|\dot \psi
   + \ii H \psi\right\rangle + \left\langle\dot \psi
   - \ii H \psi\bigg|\deriv{\psi}{\bm{ z}}\delta\dot{\bm z}\right\rangle = 0.
\end{equation}
The variational parameters $\bm z$ are complex quantities and therefore the 
variations $\delta \dot z_k$ and $\delta \dot z_k^*$ for $k=1,\dots,M$ are 
independent. 
Therefore both brackets of Eq.~\eqref{eq:braketcondif} have to vanish 
separately. 
This finally yields the equation
\begin{align}
\label{eq:lachlan_result}
 \left\langle\deriv{\psi}{\bm{z}}\bigg|\ii\dot\psi-H\psi\right\rangle = 0 \; ,
\end{align}
which is easily transformed to an implicit dynamical set of equations
$\bm K \dot{\bm z} = -\ii\bm h$ for the variational parameters with
\begin{align*}
 \bm K = \left\langle\deriv{\psi}{\bm z}\bigg|
    \deriv{\psi}{\bm z}\right\rangle \; , \quad
 \bm h = \left\langle\deriv{\psi}{\bm z}\bigg|H \psi\right\rangle \; .
\end{align*}
Up to this point there has been no specification of the 
\newpage
\noindent
trial wave function 
$\psi$ or the Hamiltonian $H$\label{todo:restrictionham}. 
The variational principle can be applied to both, linear and nonlinear 
Hamiltonians.
In the following we apply the time-dependent variational principle (TDVP)
to a trial wave function given as coupled Gaussian wave functions, and derive
dynamical equations for the variational parameters of each Gaussian.

\subsection{Dynamical equations for condensates without symmetries}
\label{subsec:eqmxyz}
We choose a superposition of $N$ Gaussians as trial wave function,
\begin{equation}
  \psi (\bm{r},t) = \sum_{k=1}^N g^k = \sum_{k=1}^N
  \ee^{\ii\left(a^k_x x^2 + a^k_y y^2 + a^k_z z^2 + \gamma^k\right)} \; .
\label{eq:gk}
\end{equation}
The $a^k_\alpha$ for $\alpha \in \{x,y,z\}$ denote complex Gaussian 
``width'' parameters in the three spatial directions, and $\gamma^k$ the 
complex amplitude/phase parameters. 
The system is described by the extended GPE 
\eqref{eq:GPE_general_timeindependent} with the Hamiltonian brought to 
the scaled ``natural'' units for the respective long-range interaction, 
\begin{equation}\label{eq:Hamiltonian586}
 \hat H = -\Delta + V_{\rm{eff}}(\bm{r}) \; ,
\end{equation}
where $V_{\rm eff}(\bm r)=V_{\rm t}(\bm r)+V_{\rm sc}(\bm r)+V_{\rm lr}(\bm r)$
is the sum of the trapping, scattering and long-range potential.
We use Eq.\ \eqref{eq:lachlan_result} obtained from the TDVP for the complex 
variational parameters
\begin{equation}
 \bm z = (\gamma^1, \hdots, \gamma^N, a_x^1,\hdots, a_x^N, a_y^1,\hdots, 
  a_y^N, a_z^1,\hdots, a_z^N) \; ,
\label{eq:z_def}
\end{equation}
and first calculate the time derivative $\dot\psi$ of the coupled Gaussian 
wave function. 
The derivation carries over to time derivatives of the Gaussian width and 
amplitude/phase parameters,
\begin{align}
\label{eq:timederivxyz}
 \frac{\dd}{\dd t} \sum_{k = 1}^{N}g^{k}
 = \sum_{k = 1}^{N} \ii \left(x^2 \dot a_x^k + y^2 \dot a_y^k + z^2 \dot a_z^k +  \dot \gamma^k \right)g^{k}.
\end{align}

\noindent
Second, we apply the Laplace operator of the Hamiltonian in 
Eq.~\eqref{eq:Hamiltonian586} in Cartesian coordinates,
\begin{align}
-\Delta \psi 
&= \sum_{k = 1}^{N}  \biggl\lbrace - 2 \ii \left[a_x^k  + a_y^k  + a_z^k \right] \nonumber \\*
&+ 4 \left[\left(a_x^k\right)^2x^2 + \left(a_y^k\right)^2y^2 + (a_z^k)^2z^2\right]\biggr\rbrace  g^k,
\end{align}
and obtain the complete expression for the ket in Eq.\ 
\eqref{eq:lachlan_result},
\begin{align}
\label{eq:unsortedvv}
  \ii \dot\psi &- H \psi = \sum_{k = 1}^{N} \biggl(
  - \biggl\lbrace  x^2 \dot a_x^k + y^2 \dot a_y^k + z^2 \dot a_z^k + 
   \dot \gamma^k \biggr\rbrace  \nonumber \\
  &- \biggl\lbrace V_{\rm{eff}}(\bm{r}) - 2 \ii \left[a_x^k  + a_y^k 
 + a_z^k \right] \nonumber \\*
  &+ 4 \left[\left(a_x^k\right)^2x^2 + \left(a_y^k\right)^2y^2
 + (a_z^k)^2z^2\right]  \biggr\rbrace  \biggr) g^{k}.
\end{align}
The sorting of Eq.~\eqref{eq:unsortedvv} according to powers of the 
coordinates $x,y,z$ results in a sum of products of a polynomial of 
second order and the Gaussian $g^k$,
\begin{align}
 \ii \dot\psi - H \psi 
 =  &\sum_{k = 1}^{N}\biggl[ v_0^k 
 + \frac{1}{2}\bigg(V_{2,x}^k x^2 \nonumber \\
 &+ V_{2,y}^k y^2 + V_{2,z}^k z^2\bigg) - V_{\rm{eff}}(\bm{r})\biggr]g^k \; ,
\end{align}
with the newly defined quantities 
\begin{subequations}
\label{eq:vVdefinition}
\begin{align}
\label{eq:v0definition}
 v^k_0 &= - \dot\gamma^k + 2 \ii \left(a_x^k  + a_y^k  + a_z^k \right) \; , \\
\label{eq:V2definition}
  \frac{1}{2} V_{2,\alpha}^k &= 
  - 4 \left(a_\alpha^k\right)^2 - \dot a_\alpha^k \; ; \; \alpha \in\{x,y,z\} \; .
\end{align}
\end{subequations}
Now, we calculate the derivatives of $\psi = \sum_{l = 1}^{N} g^l$ 
in Eq.~\eqref{eq:lachlan_result} with respect to the variational
parameters $\bm z$ in Eq.~\eqref{eq:z_def} for each $l = 1,\dots,N$:
\begin{align}
 \deriv{\psi}{\gamma^l} = \ii g^l  \; , \quad
 \deriv{\psi}{a_\alpha^l} = \ii x_\alpha^2 g^l \; ; \quad
 \alpha \in \left\lbrace x,y,z \right\rbrace.
\end{align}
Finally, Eq.~\eqref{eq:lachlan_result} results in the $4N$-dimensional system 
of equations 
\begin{widetext}
\begin{equation}
\biggl<
 \eta^2  g^l
  \biggm|  
\sum_{k = 1}^{N}\left[ v_0^k + \frac{1}{2}\left(V_{2,x}^k x^2 + V_{2,y}^k y^2 + V_{2,z}^k z^2\right) - 
V_{\rm{eff}}(\bm{r})\right]g^k  
\biggr> = 0,
\end{equation}
with $l = 1,\hdots,N$ and $\eta = 1,x,y,z$, which can be sorted as
\begin{align}\label{eq:vvglsystem}
\sum_{k = 1}^{N} \bigl< g^l \bigm| g^k \bigr> v^k_0 + 
\frac{1}{2} \sum_{k = 1}^{N}\sum_{\alpha} \bigl< g^l \bigm|x_\alpha^2 \bigm| g^k \bigr> V_{2,\alpha}^k
&=  \sum_{k = 1}^{N} \bigl< g^l \bigm| V_{\rm{eff}} \bigm| g^k \bigr>; \quad l=1,\hdots,N;\nonumber \\
\sum_{k = 1}^{N} \bigl< g^l\bigm|x_\beta^2 \bigm| g^k \bigr> v^k_0 + 
\frac{1}{2} \sum_{k = 1}^{N}\sum_{\alpha} \bigl< g^l \bigm|x_\alpha^2 x_\beta^2 \bigm| g^k \bigr> V_{2,\alpha}^k
&=  \sum_{k = 1}^{N} \bigl< g^l \bigm|x_\beta^2 V_{\rm{eff}} \bigm| g^k \bigr>; \quad
\beta\in \left\lbrace x,y,z\right\rbrace;\, l=1,\hdots,N,
\end{align}
\end{widetext}
where for a BEC without symmetries $\alpha,\beta \in \{x,y,z\}$. 
Equation~\eqref{eq:vvglsystem} can be written in the form of a matrix equation
\begin{equation}\label{eq:vvglsystemmatrixform}
 \bm{M} \bm{v} = \bm{r},
\end{equation}
with the Hermitian, positive definite $4 N \times 4 N$ matrix $\bm{M}$,
\begin{equation}
\label{eq:vVmatrix}
\bm{M} =
\begin{pmatrix}
(1)_{lk} & (x^2)_{lk} & (y^2)_{lk}  &  (z^2)_{lk}   \\
( x^2)_{kl}   &  (x^4 )_{lk}     &(x^2y^2)_{lk}   & (x^2z^2 )_{lk}  \\
( y^2)_{kl}   &  (y^2x^2)_{kl}   &(y^4   )_{lk}   &(y^2z^2 )_{lk}  \\
(z^2)_{kl} &  (z^2x^2 )_{kl}  &(z^2y^2  )_{kl} & (z^4 )_{lk}
\end{pmatrix},
\end{equation}
where all terms 
are $N\times N$ matrices for $k=1,\hdots,N$ and $l=1,\dots,N$.
As an example the term $(x^2)_{lk}$ reads
\begin{equation}
 (x^2)_{lk} = 
\begin{pmatrix}
 \langle g^{l=1}|x^2|g^{k=1}\rangle&\cdots &\langle g^{l=1}|x^2|g^{k=N}\rangle \\
 \vdots & & \vdots \\
 \langle g^{l=N}|x^2|g^{k=1}\rangle& \cdots &\langle g^{l=N}|x^2|g^{k=N}\rangle 
\end{pmatrix} \; .
\end{equation}
The terms denoted $(y^2)_{lk},(z^2)_{lk},(x^2y^2)_{lk},\hdots$, 
are analog to the example and obtained 
by replacing the $x^2$ by $y^2$, $z^2$, $x^2y^2 $, $\hdots$, respectively.
The vectors $\bm{v}$ and  $\bm{r}$ in Eq.~\eqref{eq:vvglsystemmatrixform} are 
\begin{equation}
\bm{v} = 
\begin{pmatrix}
 v_0^k \\[3pt]
 \frac{1}{2} V_{2,x}^k  \\[1ex]
 \frac{1}{2} V_{2,y}^k \\[1ex]  \frac{1}{2} V_{2,z}^k
\end{pmatrix} \; ,
\end{equation}
\begin{equation}
\label{eq:vVrighthandside}
\bm{r} =  \sum_{k = 1}^{N}	
\begin{pmatrix}
\langle g^l|V_{\rm{eff}}|g^k\rangle\\[2pt]
\langle g^l|x^2 V_{\rm{eff}}|g^k\rangle\\[2pt]
\langle g^l|y^2 V_{\rm{eff}}|g^k\rangle\\[2pt]
\langle g^l|z^2 V_{\rm{eff}}|g^k\rangle\\[2pt]
\end{pmatrix} \; ,
\end{equation}
where each entry is a vector of length $N$ for $k=1,\hdots,N$ and 
$l=1,\hdots,N$ respectively.

By solving the definitions of $(v_0^k,\bm{V}_2^k)$ in Eq.~\eqref{eq:vVdefinition} 
for the time derivatives of the Gaussian parameters, we obtain $4N$ dynamical 
equations for the Gaussian parameters $\bm{z}$,
\begin{subequations}\label{eq:eomFullSymmetry12}
\begin{align}
\dot\gamma^k &= 2 \ii (a_x^k + a_y^k + a_z^k) - v_0^k  
\label{eq:eomFullSymmetry1} \\
\dot a_\alpha^k &= - 4 (a_\alpha^k)^2 - \frac{1}{2} V_{2,\alpha}^k;\quad\alpha \in \left\lbrace x,y,z\right\rbrace;~ k=1,\hdots,N
\label{eq:eomFullSymmetry2},
\end{align} 
\end{subequations}
keeping in mind that the quantities 
$(v^k_0,\bm{V}_2^k) = (v^k_0,V_{2,x}^k,V_{2,y}^k,V_{2,z}^k)$ constitute the 
solution vector to the linear set of equations~\eqref{eq:vvglsystemmatrixform}.
These linear equations contain basic Gaussian integrals in the 
matrix \eqref{eq:vVmatrix} on the left hand side, as well as integrals with 
the interaction terms of the Hamiltonian in the 
vector \eqref{eq:vVrighthandside} on the right hand side.
The necessary integrals will be calculated analytically in 
Sec.~\ref{sec:dipolar_integrals} for condensates with rather general 
long-range interactions.

\subsection{Dynamical equations for condensates with axial or spherical symmetry}
\label{sec:eqmr}
If the GPE describes a system that is constrained by, e.g., axial or 
spherical symmetry, the results obtained in Sec.~\ref{subsec:eqmxyz} 
may be adapted to the case at hand. 
Therefore we introduce respective coordinates $(\varrho, \phi, z)$ for axial 
symmetry and $(r, \theta, \phi)$ for spherical symmetry and choose suitable 
trial wave functions 
\begin{align}\label{def:trialwave_zyl_sph}
 \psi(\bm{r},t) =\sum_{k = 1}^{N} \left[ \ee^{\ii \gamma^{k}} 
 \prod_{\alpha} \left(\ee^{\ii a_\alpha^{k}x_\alpha^{2}}\right)\right] \; ,
\end{align}
where for axially symmetric BEC we have $\alpha \in \{\varrho,z \}$, 
and for spherically symmetric, e.g., monopolar BEC we have $\alpha = r$.
These trial wave functions reduce the number of complex parameters to 
$3N$ $(a^k_\varrho,a^k_z,\gamma^k)$ and $2N$ $(a^k_r,\gamma^k)$ ($k=1\hdots,N$),
respectively, where $N$ is the number of Gaussians.

The procedure is the same as in Sec.~\ref{subsec:eqmxyz}: 
First, we calculate the time derivative $\dot \psi(t) $ which can be obtained 
by Eq.~\eqref{eq:timederivxyz} by simply setting $a^k_x = a^k_y = a^k_{\varrho}$ 
and $a^k_x = a^k_y = a^k_z = a^k_{r}$, accordingly. 
Second, the Laplace operator is applied to the coupled Gaussian wave function, 
and finally with the respective definitions of the vectors 
$\left(v_0^k,\nicefrac{1}{2}V_{2,\varrho}^k,\nicefrac{1}{2}V_{2,z}^k \right)^\mathrm{T}$ 
and $\left(v_0^k,\nicefrac{1}{2}V_{2,r}^k \right)^\mathrm{T}$,
the dynamical equations can be written as
\begin{align}
\label{eq:eomReducedSymmetry}
	\dot\gamma^k &= 2 \ii (2 a_\varrho^k + a_z^k) - v_0^k \nonumber \\
        \dot a_\alpha^k &= - 4 (a_\alpha^k)^2 - \frac{1}{2} V_{2,\alpha}^k; \quad
  k=1,\hdots,N;\, \alpha\in \left\lbrace \varrho,z\right\rbrace,
\end{align}
for a BEC with axial symmetry $(D=2)$, and 
\begin{align}
\label{eq:eomReducedSymmetry_r}
 \dot\gamma^k &= 6 \ii a_r^k - v_0^k \nonumber \\
 \dot a_r^k &= - 4 (a_r^k)^2 - \frac{1}{2} V_{2,r}^k; \quad k=1,\hdots,N,
\end{align}
for a BEC with spherical symmetry $(D=1)$.
The quantities 
$\left(v_0^k,\nicefrac{1}{2}V_{2,\varrho}^k,\nicefrac{1}{2}V_{2,z}^k \right)^\mathrm{T}$ 
and $\left(v_0^k,\nicefrac{1}{2}V_{2,r}^k \right)^\mathrm{T}$ 
have to be calculated from an adapted set of linear equations analog 
to Eq.~\eqref{eq:vvglsystem}, but for a BEC with axial symmetry, $D=2$, with
\begin{equation}
 \alpha,\beta \in \left\lbrace \varrho,z \right\rbrace, 
\end{equation}
and for a BEC with spherical symmetry, $D=1$, with
\begin{equation}
 \alpha,\beta=r. 
\end{equation}

As in Eq.~\eqref{eq:vvglsystemmatrixform}, we can rewrite both respective 
linear sets of equations in the form of a matrix equations
\begin{equation}
\label{eq:vVmatrixformreduced}
 \bm{M} \bm{v} = \bm{r},
\end{equation}
with the Hermitian, positive definite $(D+1)N \times (D+1)N$ matrix $\bm{M}$ 
analog to Eq.~\eqref{eq:vVmatrix} but with a reduced number of blocks, 
$\left\lbrace(\varrho^2)_{kl},(z^2)_{kl},\hdots\right\rbrace$ or
$\left\lbrace(r^2)_{kl},\hdots\right\rbrace$,
and the $(D+1)N$-dimensional respective vectors $\bm v$ and $\bm r$ 
for $D = 2$ and $D = 1$.

We have applied the time-dependent variational principle of Dirac Frenkel 
and McLachlan to the extended Gross-Pitaevskii equation and a trial wave 
function with coupled Gaussian functions. 
With the resulting dynamical equations \eqref{eq:eomFullSymmetry12}, 
\eqref{eq:eomReducedSymmetry} and \eqref{eq:eomReducedSymmetry_r} in 
respective symmetries we can calculate the time dependence of the wave 
function $\psi$ by calculating the time dependence of the variational 
parameters.
To set up the equations \eqref{eq:eomFullSymmetry12}, 
\eqref{eq:eomReducedSymmetry} and \eqref{eq:eomReducedSymmetry_r} we have 
to solve the set of linear equations for the quantities 
$\left(v_0^k, V_{2,\alpha}^k \right)$, $k=1,\hdots,N; \alpha=x,y,z$ 
($\alpha = \varrho,z$ and $\alpha=r$ respectively)  
in Eq.~\eqref{eq:vvglsystem}. 
All integrals of the matrix and the right hand side are calculated
analytically.

\section{Calculation of the integrals}
\label{sec:dipolar_integrals}
For clarity we sort all integrals as they appear in this matrix equation: 
the integrals for the matrix \eqref{eq:vVmatrix} in 
Sec.~\ref{subsec:vVmatrix_dipolar}, and the integrals of the right hand 
side \eqref{eq:vVrighthandside} in Sec.~\ref{subsec:vVRhs_dipolar}.
Then we calculate the mean field energy and the chemical potential for a 
BEC with long-range interaction in Sec.~\ref{sec:energyFunctionaldipolar}.

\subsection{Computation of the matrix $\bm{M}$}
\label{subsec:vVmatrix_dipolar}
The integrals of the matrix $\bm{M}$ in Eq.~\eqref{eq:vVmatrix} are all of 
the form 
$\bigl< g^l\big|g^k\bigr>$,
$\bigl< g^l  \big|  x_\alpha^2\big| g^k \bigr>$, and
$\bigl< g^l  \big|  x_\alpha^2 x_\beta^2\big| g^k \bigr>$, 
with $x_\alpha,x_\beta \in \left\lbrace x,y,z \right\rbrace$, 
and each Gaussian function $g^k$ defined in Eq.~\eqref{eq:gk}.
All integrals are easily calculated from the simplest integral 
$\langle g^l|g^k\rangle$ with the use of the relation,
\begin{equation}
\label{eq:fabcictrick_xyz}
 \bigl\langle g^l \big| x_\alpha^{2\lambda} x_\beta^{2\nu} V \big| g^k \bigr\rangle
  =  (- \ii)^{\lambda+\nu} 
 \dfrac{\partial^\lambda}{\partial \big(a^k_\alpha\big)^\lambda}
 \dfrac{\partial^\nu}{\partial \big(a^k_\beta\big)^\nu}
 \bigl\langle g^l \bigl| V \bigr| g^k \bigr\rangle \; ,
\end{equation}
with $\lambda,\nu = 0,1,2,\dots$, and $V$ an arbitrary potential.

To facilitate reading and to shorten the extensive terms in integrals or 
integral solutions, we introduce the following abbreviations:\\
\begin{subequations}\label{def:abbreviations_dipolar}
\begin{align}\label{def:abbreviations}
 a_\alpha^{kl} &= a_\alpha^k - \left(a_\alpha^l\right)^*, \\[1ex]
 a_\alpha^{ij} &= a_\alpha^i - \left(a_\alpha^j\right)^*, \\[1ex]
 a_\alpha^{klij} &=  a_\alpha^{kl} + a_\alpha^{ij}, \quad \alpha\in \{x,y,z\}, \\[1ex]
 \gamma^{kl} &= \gamma^k - \left(\gamma^l\right)^*, \\[1ex]
 \gamma^{ij} &= \gamma^i - \left(\gamma^j\right)^*, \\[1ex]
 \gamma^{klij} &=  \gamma^{kl} + \gamma^{ij}.
\end{align}
\end{subequations}
We start with
\begin{align}
\label{eq:normint_dipolar}
\bigl< g^l  \big|   g^k \bigr> 
&=\ee^{\ii \gamma^{kl}}\int\limits_{-\infty}^\infty \! \ee^{\ii a_x^{kl} x^2 }\dd x
  \int\limits_{-\infty}^\infty \!	\ee^{\ii a_y^{kl} y^2 }\dd y
  \int\limits_{-\infty}^\infty \! \ee^{\ii a_z^{kl} z^2 }\dd z \nonumber \\
&=\frac{\pi^{\nicefrac{3}{2}} \ee^{\ii \gamma^{kl}}}
 {\sqrt{-\ii a^{kl}_x}\sqrt{-\ii a^{kl}_y}\sqrt{-\ii a^{kl}_z}} \; ,
\end{align}
where, as can be seen easily from the definition of the Gaussian trial 
wave function \eqref{def:dipolartrialfunction}, the imaginary parts of the 
widths are to remain positive. 
Therefore the imaginary parts of the occurring combinations also fulfill 
$\mathrm{Im}\,[a^k - a^{l*}]>0 $.

The application of the relation \eqref{eq:fabcictrick_xyz} to the norm integral 
\eqref{eq:normint_dipolar} provides with $\lambda=1$, $\nu=0$ the integrals
\begin{subequations}
\begin{align}
\label{eq:dipolarmatrixfirst}
 \bigl< g^l  \bigm| x^2 \bigm|  g^k \bigr> &= 
 \frac{\pi^{\nicefrac{3}{2}} \ee^{\ii \gamma^{kl}}} 
{2 \left(-\ii a^{kl}_x\right)^{\nicefrac{3}{2}} \sqrt{-\ii a^{kl}_y}\sqrt{-\ii a^{kl}_z}},\\
\bigl< g^l  \bigm| y^2 \bigm|  g^k \bigr> &= 
 \frac{\pi^{\nicefrac{3}{2}} \ee^{\ii \gamma^{kl}}} 
{2\sqrt{-\ii a^{kl}_x}\left(-\ii a^{kl}_y\right)^{\nicefrac{3}{2}}\sqrt{-\ii a^{kl}_z}},\\
\bigl< g^l  \bigm| z^2 \bigm|  g^k \bigr> &= 
 \frac{\pi^{\nicefrac{3}{2}} \ee^{\ii \gamma^{kl}}} 
{2\sqrt{-\ii a^{kl}_x}\sqrt{-\ii a^{kl}_y}\left(-\ii a^{kl}_z\right)^{\nicefrac{3}{2}}},
\end{align}
\end{subequations}
and with $\lambda+\nu=2$ the integrals
\begin{subequations}
\begin{align}
  \bigl< g^l  \bigm| x^4 \bigm|  g^k \bigr> &= 
 \frac{3\pi^{\nicefrac{3}{2}} \ee^{\ii \gamma^{kl}}} 
{4 \left(-\ii a^{kl}_x\right)^{\nicefrac{5}{2}} \sqrt{-\ii a^{kl}_y}\sqrt{-\ii a^{kl}_z}},\\
 \bigl< g^l  \bigm| y^4 \bigm|  g^k \bigr> &= 
 \frac{3\pi^{\nicefrac{3}{2}} \ee^{\ii \gamma^{kl}}} 
{4\sqrt{-\ii a^{kl}_x}\left(-\ii a^{kl}_y\right)^{\nicefrac{5}{2}}\sqrt{-\ii a^{kl}_z}},\\
\bigl< g^l  \bigm| z^4 \bigm|  g^k \bigr> &= 
 \frac{3\pi^{\nicefrac{3}{2}} \ee^{\ii \gamma^{kl}}} 
{4\sqrt{-\ii a^{kl}_x}\sqrt{-\ii a^{kl}_y}\left(-\ii a^{kl}_z\right)^{\nicefrac{5}{2}}},
\end{align}
\label{eq:x4terms}
\end{subequations}
and
\begin{subequations}
\begin{align}
  \bigl< g^l  \bigm| x^2y^2 \bigm|  g^k \bigr> &= 
 \frac{\pi^{\nicefrac{3}{2}}\ee^{\ii \gamma^{kl}}} 
{4 \left(-\ii a^{kl}_x\right)^{\nicefrac{3}{2}} \left(-\ii a^{kl}_y\right)^{\nicefrac{3}{2}}\sqrt{-\ii a^{kl}_z}},\\
\bigl< g^l  \bigm| x^2z^2 \bigm|  g^k \bigr> &= 
 \frac{\pi^{\nicefrac{3}{2}}\ee^{\ii \gamma^{kl}}} 
{4 \left(-\ii a^{kl}_x\right)^{\nicefrac{3}{2}} \sqrt{-\ii a^{kl}_y}\left(-\ii a^{kl}_z\right)^{\nicefrac{3}{2}}},\\
\label{eq:dipolarmatrixlast}
\bigl< g^l  \bigm| y^2z^2 \bigm|  g^k \bigr> &= 
 \frac{\pi^{\nicefrac{3}{2}}\ee^{\ii \gamma^{kl}}} 
{4\sqrt{-\ii a^{kl}_x}\left(-\ii a^{kl}_y\right)^{\nicefrac{3}{2}}\left(-\ii a^{kl}_z\right)^{\nicefrac{3}{2}}}.
\end{align}
\label{eq:x2y2terms}
\end{subequations}
Since the matrix is now complete, we turn to the more challenging integrals 
needed for the right hand side of Eq.~\eqref{eq:vvglsystemmatrixform}, i.e.,
the vector $\bm r$ defined in  Eq.~\eqref{eq:vVrighthandside}.

\subsection{Computation of the vector $\bm{r}$}
\label{subsec:vVRhs_dipolar}
The right hand side of Eq.~\eqref{eq:vvglsystemmatrixform} contains the trapping term, $V_\mathrm{t}$, 
the scattering term, $V_\scc$, 
as well as the more complicated long-range interaction term, $V_\mathrm{lr}$. 
In the following equations the $g^k$ defined in Eq.~\eqref{eq:gk}
represent the individual Gaussians of the trial wave function.

\subsubsection{Trapping}
\label{subsubsec:trap}
We start with the term of the trapping potential
\begin{align}\label{eq:dipolar_rhs_traps}
&\bigl< g^l  \bigm| V_\mathrm{t}  \bigm|  g^k \bigr> = 
\bigl< g^l  \bigm| \gamma_x^2 \, x^2 + \gamma_y^2 \, y^2 + \gamma_z^2 \, z^2 \bigm|  g^k \bigr> \nonumber \\
&=\frac{
\pithree \ee^{\ii \gamma^{kl}}
	\left( \gamma_x^2 a_y^{kl}a_z^{kl}
	      + \gamma_y^2 a_x^{kl}a_z^{kl}
	      + \gamma_z^2 a_x^{kl}a_y^{kl}
	\right)}
  {2 (-i a_x^{kl})^{3/2} (-i a_y^{kl})^{3/2} (-i a_z^{kl})^{3/2}}.
\end{align}
For the right hand side of Eq.~\eqref{eq:vvglsystemmatrixform} we also need 
the terms
\begin{align*}
 \bigl< g^l  \bigm| x_\alpha^{2} V_\mathrm{t} \bigm|  g^k \bigr>,
\end{align*}
with $x_\alpha \in  \left\lbrace x,y,z \right\rbrace $.
They are directly obtained with the help of Eqs.\ \eqref{eq:x4terms}
and \eqref{eq:x2y2terms}.

\subsubsection{Scattering}
The second interaction term of $V_\mathrm{eff} $ on the right hand side of 
Eq.~\eqref{eq:vvglsystemmatrixform} contains the nonlinear contact interaction
$V_\scc = 8\pi\scatt \left|\psi(\bm r) \right|^2$ of the s-wave scattering.
Following the same procedure as above, we start by calculating 
$ \bigl< g^l  \big|  V_\scc  \big|  g^k \bigr>$, before obtaining the 
terms $ \bigl< g^l  \big| x_\alpha^2 V_\scc  \big|  g^k \bigr>$ 
with $x_\alpha \in \left\lbrace x,y,z \right\rbrace$ using the 
relation \eqref{eq:fabcictrick_xyz}.
\begin{align}
\label{eq:dipolar_rhs_scatter}
  &\bigl< g^l  \bigm|  V_\scc  \bigm|  g^k \bigr> =
\bigl< g^l  \bigm| 8 \pi \scatt \left|\psi \right|^2  \bigm|  g^k \bigr> \nonumber  \\
&= \sum_{i,j=1}^N  \int\dd^3\bm r\, \left( 8 \pi \scatt \left.g^l (\bm r)\right.^* \left.g^j(\bm r)\right.^* g^k(\bm r) g^i(\bm r) \right)
 \nonumber  \\
&= 8 \scatt \pifive \sum_{i,j=1}^N \frac{\ee^{\ii{\gamma^{klij}}}}
{\sqrt{-\ii a^{klij}_x}\sqrt{-\ii a^{klij}_y}\sqrt{-\ii a^{klij}_z}},
\end{align}
as long as $\mathrm{Im}\, a^{klij}_\alpha >0$ for 
$\alpha \in \left\lbrace x,y,z \right\rbrace $, which is true 
[see Eq.~\eqref{def:abbreviations_dipolar} and note that 
$\mathrm{Im}\, a^{k}_\alpha>0$ for all width parameters].
Again we use relation \eqref{eq:fabcictrick_xyz} and get the three 
remaining integrals,
\begin{subequations}
\begin{align}
   &\bigl< g^l  \bigm| x^2 V_\scc  \bigm|  g^k \bigr> \nonumber \\* &=
 4 \scatt \pifive \sum_{i,j=1}^N \frac{\ee^{\ii{\gamma^{klij}}}}
{\left(-\ii a^{klij}_x\right)^{\nicefrac{3}{2}}\sqrt{-\ii a^{klij}_y}\sqrt{-\ii a^{klij}_z}}, \\
 &\bigl< g^l  \bigm| y^2 V_\scc  \bigm|  g^k \bigr> \nonumber \\*&= 
 4 \scatt \pifive \sum_{i,j=1}^N \frac{\ee^{\ii{\gamma^{klij}}}}
{\sqrt{-\ii a^{klij}_x}\left(-\ii a^{klij}_y\right)^{\nicefrac{3}{2}}\sqrt{-\ii a^{klij}_z}}, \\
 \label{eq:dipolar_rhs_scatterz2}
 &\bigl< g^l  \bigm| z^2 V_\scc  \bigm|  g^k \bigr> \nonumber \\*&=
 4 \scatt \pifive \sum_{i,j=1}^N \frac{\ee^{\ii{\gamma^{klij}}}}
{\sqrt{-\ii a^{klij}_x}\sqrt{-\ii a^{klij}_y}\left(-\ii a^{klij}_z\right)^{\nicefrac{3}{2}}}.
\end{align}
\end{subequations}

\subsubsection{Long-range interaction}
\label{subsubsec:long-range}
The most challenging calculation surely is that of the long-range 
interaction term (see Eq.~\eqref{eq:GPE_general_timeindependent})
$V_\mathrm{lr} =  \int \mathrm{d}^3 \bm{r}' 
W_{\mathrm{lr}}\!\left( \bm r - \bm r' \right)
\left | \psi(\bm{r}')\right |^2 $,	
\begin{align}
\label{eq:dipolar_integral}
  \bigl< g^l  \bigm| V_\mathrm{lr}  \bigm|  g^k \bigr>  = &\sum^N_{i,j=1}
\int \! \mathrm{d}^3 \bm{r} \int \! \mathrm{d}^3 \bm{r}' \, 
 W_{\mathrm{lr}}\!\left( \bm r - \bm r' \right)  \nonumber \\*
&\times \left. g^j \right.^*(\bm r')  g^i(\bm r')  \left. g^l \right.^*(\bm r)  g^k(\bm r).
\end{align}
Introducing relative (rel) and ``center of mass'' coordinates (cm) via
\begin{align}
\begin{pmatrix}\bm r\\\bm r' \end{pmatrix} &= \frac{1}{2}
\begin{pmatrix}1 & 1 \\-1 & 1 \end{pmatrix}
\begin{pmatrix}\bm r_\mathrm{rel}\\\bm r_\mathrm{cm} \end{pmatrix},
\end{align}
and keeping the Jacobian determinant (\nicefrac{1}{8}) in mind, 
Eq.~\eqref{eq:dipolar_integral} is transformed to
\begin{align}\label{eq:rcm_integration}
&\bigl< g^l  \bigm| V_\mathrm{lr}  \bigm|  g^k \bigr> 
= 
 \sum^N_{i,j=1} \frac{\ee^{\ii{\gamma^{klij}}}}{8}
\int \! \mathrm{d}^3 \bm{r}_\mathrm{rel} \int \! \mathrm{d}^3 \bm{r}_\mathrm{cm} \,
W_{\mathrm{lr}}\!\left( \bm{r}_\mathrm{rel} \right)   \nonumber \\
&\, \times \mathrm{exp}\Biggl\lbrace \frac{\ii}{4} \Bigl( a_x^{kl} \left(x_\mathrm{cm} + x_\mathrm{rel}\right)^2
	     + a_y^{kl} \left(y_\mathrm{cm} + y_\mathrm{rel}\right)^2   \nonumber \\*
		&\, \qquad \quad + a_z^{kl} \left(z_\mathrm{cm} + z_\mathrm{rel}\right)^2
  +a_x^{ij} \left(x_\mathrm{cm} - x_\mathrm{rel}\right)^2   \nonumber \\*
	     &\, \qquad \quad+ a_y^{ij} \left(y_\mathrm{cm} - y_\mathrm{rel}\right)^2
		+ a_z^{ij} \left(z_\mathrm{cm} - z_\mathrm{rel}\right)^2
\Bigr)\Biggr\rbrace,
\end{align}
with the abbreviations \eqref{def:abbreviations_dipolar}.
We use the integral 
\begin{align}\label{eq:hilfsintegral123}
\int\limits_{-\infty}^{\infty}\ee^{\ii\left({a(x+u)^2+b(x-u)^2}\right)}\dd x 
 = \frac{\sqrt{\pi}}{\sqrt{-\ii(a+b)}}\ee^{\ii\frac{ 4 a b u^2}{a+b}}
\end{align}
which, for $\mathrm{Im}[a+b]>0$, is easily transformed into a standard 
Gaussian integral after completing the square in the exponent, 
and solve the center of mass integral in Eq.~\eqref{eq:rcm_integration}
\begin{align}\label{eq:anschlussruediger}
\bigl< g^l  \bigm| V_\mathrm{lr}  \bigm|  g^k \bigr> 
&=
 \sum^N_{i,j=1}  \pithree  \frac{\ee^{\ii{\gamma^{klij}}}}{\sqrt{\ii a_x^{klij} a_y^{klij} a_z^{klij}}}~~
\mathrm{I}^{klij}_\mathrm{rel},
\end{align}
with
\begin{align}
\label{eq:dipolarIrel}
 &\mathrm{I}^{klij}_\mathrm{rel} = 
 \int \! \mathrm{d}^3 \bm{r}_\mathrm{rel} \Biggl[
 W_{\mathrm{lr}}\!\left( \bm{r}_\mathrm{rel} \right) \Biggl. \nonumber \\
&\times \Biggl. \mathrm{exp}\left\lbrace\ii \left(
\frac{a_x^{ij}a_x^{kl}}{a_x^{klij}}x_\mathrm{rel}^2 + 
\frac{a_y^{ij}a_y^{kl}}{a_y^{klij}}y_\mathrm{rel}^2 +
\frac{a_z^{ij}a_z^{kl}}{a_z^{klij}}z_\mathrm{rel}^2 
\right)\right\rbrace \Biggr].
\end{align}
The relative integral $\mathrm{I}^{klij}_\mathrm{rel}$ depends on the 
specific form of the two-particle interatomic interaction 
$W_{\mathrm{lr}}\!\left( \bm{r}_\mathrm{rel} \right)$. 
The integral obtained with the coupled Gaussian method is formally the same 
as for the calculation with a single Gaussian trial wave function, except for 
the complex coefficients in the exponent,
 $a_\alpha^{ij}a_\alpha^{kl}/a_\alpha^{klij}$ 
for $\alpha \in \left\lbrace x,y,z\right\rbrace$.  
Therefore the method of coupled Gaussian functions is applicable to all 
two-particle long-range interactions which can be solved with the 
simple single Gaussian wave function.

As an example, we present the results for the relative integral for 
dipolar interaction
\begin{equation}
 W_{\mathrm{lr}}\!\left( \bm{r}_\mathrm{rel} \right) = 
 W_{\mathrm{d}}\!\left( \bm{r}_\mathrm{rel} \right) =
  \frac{1-3\frac{z_\mathrm{rel}^2}{|\bm r_\mathrm{rel}|^2}} {\left | \bm{r}_\mathrm{rel} \right |^3},
\end{equation}
where $I_\mathrm{rel}$ can be expressed in terms of 
elliptic integrals \cite{Tikhonenkov08a,ruedigerDA},
\begin{align}
\label{eq:dipolarIrel_geklaut}
\mathrm{I}_\mathrm{rel} = \frac{4 \pi}{3} \left[ \kappa_x \kappa_y R_D\left( \kappa_x^2,\kappa_y^2,1 \right)   - 1 \right] \; .
\end{align}
$\kappa_x,\kappa_y$ are complex combinations of the Gaussian widths,
\begin{align*} 
 \kappa_x = \sqrt{\frac{a_x^{klij}a_z^{ij}a_z^{kl}}
{a_x^{ij}a_x^{kl}a_z^{klij}} } \; , \quad
\kappa_y = \sqrt{\frac{a_y^{klij}a_z^{ij}a_z^{kl}}
{a_y^{ij}a_y^{kl}a_z^{klij}} }, 
\end{align*}
and $R_D$ is the elliptic integral of the second kind in 
Carlson form \cite{Carlson95a,Carlson94a},
\begin{equation*}\label{def:RDCarlson}
 R_D(x,y,z) = \frac{3}{2} \int\limits_0^\infty \! \frac{\mathrm{d}t}
{\sqrt{(x+t)(y+t)(z+t)^3}} \; .
\end{equation*}
Numerically it is convenient to use Carlson's formulation for 
elliptic integrals because there are very fast converging algorithms 
available \cite{Carlson95a,Carlson94a} even for complex arguments 
$x,y,z \in \mathbb{C}$.

The three additional integrals 
$\bigl< g^l  \big| x_\alpha^2 V_\mathrm{lr}  \big|  g^k \bigr>$
for $ x_\alpha \in \left\lbrace x,y,z \right\rbrace$
needed to complete Eq.~\eqref{eq:vvglsystemmatrixform}
are obtained using derivatives of $\langle g^l|V_\mathrm{lr}|g^k\rangle$
with respect to the Gaussian widths $a_\alpha^k $, 
see Eq.~\eqref{eq:fabcictrick_xyz},
\begin{align*}
 \bigl< g^l  \bigm| x_\alpha^2 V_\mathrm{lr}  \bigm|  g^k \bigr> 
 =  -\ii\deriv{}{a_\alpha^k}\bigl< g^l \bigm| V_\mathrm{lr} \bigm| g^k \bigr> \; .
\end{align*}

\subsection{Energy functional and chemical potential}
\label{sec:energyFunctionaldipolar}
We calculate the mean field energy and the chemical potential,
\begin{subequations}
\begin{align}
\label{eq:energyfunctional_dipolar} 
 E_{\rm mf} &= \sum_{k,l=1}^N \bigl< g^l \bigm| -\Delta
 + V_\mathrm{t}+\frac{1}{2}\left(V_\scc+V_\mathrm{lr}\right)\bigm| g^k\bigr> \; ,\\
  \mu &= \sum_{k,l=1}^N \langle g^l| -\Delta + V_{\rm t}
     + V_\scc+V_{\rm lr}|g^k\rangle \; .
\label{eq:mufunctional_dipolar}
\end{align}
\end{subequations}
The terms for trapping, scattering, and the long-range interaction have 
been evaluated in Sec.~\ref{subsec:vVRhs_dipolar}.
We now calculate the kinetic term and apply the Laplace operator to 
the coupled Gaussian wave 
function,
\begin{align}
\langle g^l|\Delta| g^k\rangle&= 
\frac{ -2 \pithree  \ee^{\ii \gamma^{kl}}}{\left(\ii a_x^{kl}a_y^{kl}a_z^{kl} \right)^{\nicefrac{1}{2}}} 
\sum_{\alpha=x,y,z} \left\lbrace \frac{ \left(a_\alpha^{k}\right)^2 }{-\ii a_\alpha^{kl} }- \ii a_\alpha^k \right\rbrace \; .
\label{eq:dipolarMeanFieldKinetic}
\end{align}

Now, all integrals that are needed for the linear equations for the 
quantities $\left(v_0^k, V_{2,\alpha}^k \right)$, $k=1,\hdots,N; \alpha=x,y,z$ 
in Eq.~\eqref{eq:vvglsystem}, as well as all integrals for the mean field 
energy and the chemical potential have been calculated.
We are able to set up the dynamical equations for the Gaussian parameters.
These dynamical equations can be solved using three different methods, 
either by minimization of the mean field energy, by the search for fixed 
points of the dynamical equations, or by imaginary time evolution of an 
initial wave function.

\section{Computation of the ground state and excited states}
\label{sec:numericalprocedure}
There are three different methods available to calculate variational solutions. 
The solutions obtained via minimization of the mean field energy 
or via evolution of an initial wave function in imaginary time, are limited 
to the stable ground state. 
The method of a highly nonlinear root search of the dynamical equations 
\eqref{eq:eomFullSymmetry12} yields all stationary states of the GPE, 
the ground state and collectively excited states.

The sensitivity of the methods on the initial values greatly differs.
While the minimization and the imaginary time evolution are relatively robust, 
the root search requires sufficiently accurate parameters, 
especially for an increasing number of variational parameters.
For more coupled Gaussian functions, 
however, the minimization of the mean field energy 
and the imaginary time evolution get increasingly time-consuming.
Results should therefore be obtained with the nonlinear root search, the 
other two routines should only be used to calculate appropriate initial values.

\subsection{Minimization of the mean field energy}
\label{sec:num_theory_meanfieldmin}
One method of obtaining the ground state of the Gross-Pitaevskii equation
is to minimize the mean field energy functional in Eq.~\eqref{eq:mf_functional_theory}.
For multiple coupled Gaussians, the analytical calculation of all derivatives 
with respect to the variational parameters, e.g., for calculating the 
gradient, is not possible. 
Therefore we use a numerical minimization routine that uses the energy 
function values only. 
For an increased number of Gaussian parameters, the accuracy of this method 
is limited, but the results may be used as initial values for a root search 
of the dynamical equations, which provides much more reliable and accurate 
results.

\subsection{Root search for fixed points of the dynamical equations}
\label{sec:num_theory_rootsearch}
The full three-dimensional calculation for condensates with long-range 
interaction includes $4 N$ complex variational parameters of the wave function.
We search for these solutions of the nonlinear dynamical equations 
\eqref{eq:eomFullSymmetry12}, \eqref{eq:eomReducedSymmetry}, and
\eqref{eq:eomReducedSymmetry_r}  where the dynamic is trivial, i.e., 
the stationary states. 
The phase of all Gaussians is defined by the chemical potential 
(see Eq.~\eqref{eq:numproc_eqm}).

We use a wave function with $N$ complex 
amplitude and phase parameters $\gamma^k,\hdots,\gamma^N$.
Since the wave function has to be normalized, we have to ensure 
$\langle \psi | \psi\rangle = 1$.
In summary, we have to find roots of the following system of equations 
\begin{align} \label{eq:numproc_eqm}
  \dot  a_x^k &= - 4 (a_x^k)^2 - \frac{1}{2} V_{2,x}^k  \stackrel{!}{=} 0 \, ,  \nonumber \\
\dot  a_y^k &= - 4 (a_y^k)^2 - \frac{1}{2} V_{2,y}^k \stackrel{!}{=} 0 \, , \nonumber \\
\dot  a_z^k &= - 4 (a_z^k)^2 - \frac{1}{2} V_{2,z}^k \stackrel{!}{=} 0 \, , \nonumber \\
\dot \gamma^k &= 2 \ii (a_x^k + a_y^k + a_z^k) - v_0^k  \stackrel{!}{=} - \mu \, , \nonumber \\
&\ \quad\left\langle \psi \mid \psi \right\rangle - 1 \stackrel{!}{=} 0 \, ,
\end{align}
for $k=1,\hdots,N$.
The quantities $v_0$ and $V_{2,x,y,z}$ constitute the solution vector to the 
set of equations~\eqref{eq:vvglsystem}. 
The right hand side \eqref{eq:vVrighthandside} of this set of equations 
contains the integrals of contact, trap and long-range interaction.
The calculation of the dipolar interaction yields elliptic integrals which 
are evaluated with the help of fast converging 
algorithms \cite{Carlson95a,Carlson94a}.

The root search itself is highly nonlinear and may be performed with an 
algorithm based on the Powell hybrid method \cite{NumericalRecipes}. 
The success and the speed of any numerical root search greatly depends on 
the number of variational parameters. 
Therefore we reduce the number of parameters prior to 
this routine as much as possible. 
Since the ground state of e.g., dipolar condensates in an axially symmetric 
trap is axially symmetric, $a^k_x = a^k_y \equiv a^k_\varrho$,
it is possible to reduce the first $3N$ equations 
in Eqs.~\eqref{eq:numproc_eqm} to $2N$ equations.
For condensates with $1/r$ interaction, the wave function as well as the 
stationary states are spherically symmetric. 
Therefore, we are able to further reduce the number of width parameters 
using $a^k_x = a^k_y = a^k_z \equiv a^k_r$.

\subsection{Imaginary time evolution}
\label{imaginary time evolution}
The third method for finding solutions of the GPE is 
the imaginary time evolution of an initial wave function. 
Although the calculation for a distinct scattering length may take a long time, 
the routine is rather insensitive to the choice of the initial parameter values,
even for a large number of variational parameters. 
Therefore, the imaginary time evolution is very useful in the context of 
the calculation of very accurate input values for the root search.

We substitute $t \rightarrow \tau = \ii t$ in the GPE and calculate the 
(imaginary) time evolution of the dynamical equations.
For the linear Schr\"odinger equation this leads to a damping of all states 
with a factor according to their respective energy eigenvalues 
$\mathrm{exp}(-E_n \tau)$.
Sufficiently long integration with respect to the imaginary time coordinate 
and renormalizing yields the ground state of the Hamiltonian. 
The method can also be applied to the nonlinear Gross-Pitaevskii equation.

In a next step, we linearize the dynamical equations in the vicinity of 
the fixed points in order to analyze the stability and possible bifurcations. 
We can also calculate fluctuations $\delta \psi$ of the stationary wave 
function.

\section{Stability Properties of the Variational Fixed Points}
\label{sec:theory:EW}
The standard method for analyzing the stability of solutions of the GPE 
is to perturb the wave function of the solution and linearize the GPE, which 
leads to the Bogoliubov-de Gennes equations \cite{PitaevskiiBlackBook}. 
In this section, we make a different approach and apply methods
of nonlinear dynamics to analyze the stability of the condensates.

The application of the TDVP to a wave function consisting of coupled Gaussian 
functions in Sec.~\ref{sec:lachlan_theory} led to a set of dynamical equations.
In Sec.~\ref{sec:numericalprocedure} we described methods for searching for 
stationary solutions of the equations \eqref{eq:numproc_eqm}. 
Minimization of the mean field energy, imaginary time evolution or root search 
yield variational parameters $\bm{z}^\mathrm{FP}$ for stationary solutions of 
the GPE, so called ``fixed points'' (FP).
In the case of $N$ coupled Gaussian functions these parameters are $3N$ 
complex widths, $N$ for each $a_x^k, a_x^k, a_z^k$ and $N$ complex 
amplitudes/phases $\gamma^k$:
\begin{align}
 \bm{z}^\mathrm{FP} = (\gamma^k, a_x^k, a_y^k, a_z^k)^\mathrm{FP}; \quad k=1,\hdots,N.
\end{align}
In the case of an axially or spherically symmetric BEC, the procedure that 
follows is the same, but with reduced sets of subscripts, $(\varrho,z)$ or 
even $(r)$ for the width parameters.

We will investigate the stability of these stationary solutions with the 
help of a perturbation of the parameters at the fixed point.
If the parameter set for the stationary solution is indeed the minimum of 
the mean field energy, we expect the solution to be stable.
In this case small changes of the variational parameters will only result 
in a quasi-periodic oscillation confined to the vicinity of the original 
fixed point.
By contrast, as we will see, if the stationary fixed point is a 
hyperbolic fixed point, small perturbations will lead to an exponential 
growth of the solution.
For the following calculation of the stability matrix, it is irrelevant 
which method was used to obtain the parameters of the stationary solution 
in the first place.

To observe the time dependence of the perturbations, we use the dynamical 
equations for the Gaussian parameters \eqref{eq:eomFullSymmetry12}.
The fixed point obviously fulfills Eqs.~\eqref{eq:numproc_eqm},
\begin{align}
\dot\gamma^k (\bm{z}^{\mathrm{FP}}) &= -\mu, \nonumber \\
\dot a_\alpha^k(\bm{z}^{\mathrm{FP}}) &= 0,
\end{align}
for $\alpha=x,y,z$; $k=1,\hdots,N$.
For small deviations from the fixed point, we first split the equations 
above into real ($\mathrm{Re}$) and imaginary ($\mathrm{Im}$) parts 
(indicated with the tilde $\bm z \rightarrow \bm {\tilde z}$) in order 
to linearize them,
\begin{equation}
\label{eq:lineom_theory}
 \delta \dot{\tilde{\bm z}} = \bm {J} \delta\tilde{\bm z}.
\end{equation}
$\delta\tilde{\bm z}$ denotes the deviation of the variational parameters 
from those at the fixed point, 
$\tilde{\bm z} = \tilde{\bm z}^{\mathrm{FP}} + \delta\tilde{\bm z}$, 
and $\bm{J}$ denotes the $8N \times 8N$-dimensional real valued Jacobian 
matrix at the fixed point
\begin{equation}\label{eq:theory_Jacobian_EW}
 \bm{J} = \dfrac{ \partial \left(\gamma^{k,\mathrm{Re}},\gamma^{k,\mathrm{Im}}
			     , a_\alpha^{k,\mathrm{Re}}, a_\alpha^{k,\mathrm{Im}} \right)}
		{ \partial \left(\gamma^{l,\mathrm{Re}},\gamma^{l,\mathrm{Im}}
			     , a_\beta^{l,\mathrm{Re}}, a_\beta^{l,\mathrm{Im}} \right)},
\end{equation}
with $\alpha,\beta = x,y,z$ and $k,l = 1,\hdots,N$. 
The eigenvalues of $\bm{J}$ determine the characteristic stability of 
the fixed point in whose surroundings the linearization takes place. 
In the coordinates given by the eigenvectors of $\bm J$, all differential 
equations take the form 
\begin{equation}\label{eq:lineom_diag_theory}
 \delta\dot{\tilde{\bm z}}_i = \lambda_i \, \delta \tilde{\bm z}_i; \quad i = 1,\hdots,8N,
\end{equation}
which have the simple solution 
\begin{equation}
   \delta \tilde{\bm z}_i (t)
 = \delta \tilde{\bm z}_i^0\mathrm{e}^{\lambda_i t}. 
\label{eq:zetapunkt}
\end{equation}
The eigenvalues occur in pairs, i.e., if $\lambda_i$ is an eigenvalue, 
$-\lambda_i$ is also an eigenvalue of $\bm J$.
Since the Jacobian matrix is not symmetric, the eigenvalues $\lambda_i$ 
are real or complex, and there are two possibilities.
If all eigenvalues $\lambda_i$ are purely imaginary the time evolution of 
the perturbation remains confined, since the solution \eqref{eq:zetapunkt}
is oscillating.
In contrast, if at least one of the real parts of the eigenvalues 
($\mathrm{Re}\,\lambda_i$) is non-zero, any perturbation in the direction 
of the corresponding eigenvector will grow exponentially.

Using this method we observe the behavior under small perturbations.
We are able to investigate the stability of any fixed point 
that we obtain for example from a root search of the dynamical equations.

For unstable fixed points the methods of nonlinear dynamics also allow us 
to gain insight in the collapse mechanism by analyzing variations of the 
wave function characterized by the respective eigenvectors
\begin{align}
\label{eq:eigenvectors_stabanal}
 \delta \tilde{\bm z}_i
 = \left(
\delta \gamma^{k,\mathrm{Re}}_i, 
\delta \gamma^{k,\mathrm{Im}}_i, 
\delta a_{\alpha,i}^{k,\mathrm{Re}},
\delta a_{\alpha,i}^{k,\mathrm{Im}}\right)^{\mathrm{T}}, 
\end{align}
with $k = 1,\hdots,N$; $\alpha \in \left\lbrace x,y,z \right\rbrace$.
If this eigenvector $\delta\tilde{\bm z}_i$ is axially symmetric, 
i.e., $\delta a_{x,i}^k = \delta a_{y,i}^k $ for all $k$, 
the perturbation of the condensate is symmetric.
If, however, the eigenvector breaks the axial symmetry, i.e., 
$\delta a_{x,i}^k = -\delta a_{y,i}^k $ for all $k$, the perturbation leads 
to an asymmetric oscillation or collapse of the condensate, depending on 
whether the respective eigenvalue is purely real or imaginary.

With the variations of the variational parameters of the eigenvector
we calculate the expansion of the wave function at the fixed point and
get the solution for the linearized variation of the wave function
\begin{align}
 \delta \psi_i(\bm r,t) = \sum_{k = 1}^{N}
\Biggl( \sum_{\alpha}\left[ \ii x_\alpha^2 \delta a_{\alpha,i}^{k,\mathrm{Re}} 
- x_\alpha^2 \delta a_{\alpha,i}^{k,\mathrm{Im}}\right] \nonumber \\
+ \ii \delta \gamma_i^{k,\mathrm{Re}} - \delta \gamma_i^{k,\mathrm{Im}} \Biggr)
g^k|^{\mathrm{FP}}(\bm r)\, \ee^{\lambda_i t},
\end{align}
with $\alpha \in \left\lbrace x,y,z \right\rbrace$.
The respective eigenvalue is denoted  $\lambda_i$ and $g^k|^{\mathrm{FP}}$ 
is the unperturbed Gaussian $k$ at the fixed point.

Is the approach based on the stability eigenvalues of the Jacobian fully
equivalent to solutions of the Bogoliubov-de Gennes equations 
\cite{PitaevskiiBlackBook}?
Probably not.
Although the ansatz in Eq.\ \eqref{def:dipolartrialfunction} is quite 
general and not restricted to spherical or axial symmetric condensates 
it allows for the description of the subset of excitations which are 
consistent with the ansatz \eqref{def:dipolartrialfunction}.
To obtain all solutions of the Bogoliubov-de Gennes equations with a
variational approach the ansatz must be further generalized.
This will be the objective of future work.

\section{Conclusion}
\label{sec:conclusion}
We used an ansatz with several coupled Gaussian functions to obtain an 
improved variational description of the dynamics of Bose-Einstein condensates.
We applied a time-dependent variational principle to the Gross-Pitaevskii 
equation and obtained dynamical equations for the variational parameters
with the improved variational method of coupled Gaussian functions. 
We discussed methods for solving these equations and analyzing the stability 
using methods of nonlinear dynamics.
When can we expect the proposed variational methods to be better than 
a grid method?
This is hard to say in general.
The variational ansatz with a finite number of Gaussians is still an
approximation.
However, grid calculations are only ``exact'' in the limit of a small
step size and thus an infinitely large grid.

For a Gross-Pitaevskii equation with long-range monopolar $(1/r)$ and
dipolar $(1/r^3)$ interaction, the improved variational results, 
the convergence, and comparisons with numerical calculations will be 
subject of the subsequent paper \cite{paper2}.
It will be shown that a low number of Gaussians is sufficient to obtain 
converged results and thus the number of parameters in the variational
computations is significantly smaller compared to calculations on grids,
and that a wealth of new phenomena is obtained by using the ansatz with
coupled Gaussian functions.

%

\end{document}